# Noise Activated Fast Locomotion of DNA through Frictional Landscape of Nanoporous Gel

*Aniruddha Deb[1], Prerona Gogoi[1], Sunil K. Singh[1], Partho Sarathi Gooh Pattader[1,2,3]\**

[1]Department of Chemical Engineering, Indian Institute of Technology Guwahati; Assam, 781039, India.
[2]Centre for Nanotechnology, Indian Institute of Technology Guwahati; Assam, 781039, India.
[3]Jyoti and Bhupat Mehta School of Health Science & Technology, Indian Institute of Technology Guwahati; Assam, 781039, India.

*Corresponding author. Email: psgp@iitg.ac.in

**ABSTRACT**. It is hypothesized that nonlinear solid friction between the gel matrix and DNA molecules inhibits the motion of DNA through the nanopores of the gel during electrophoresis. In this article, it is demonstrated that external noise can alleviate the effect of solid friction, thus enhancing the mobility of DNA in an electrophoretic setting. In the presence of noise, the mobility of DNA increases by more than ~113 % compared to conventional electrophoresis. Although at a high power of noise, DNA exhibits Arrhenius kinetics, at a low power of noise, super Arrhenius kinetics suggest the collective behavior of the activated motion of DNA molecules. Stochastic simulation following modified Langevin dynamics with the asymmetric pore size distribution of the agarose gel successfully predicts the mobility of DNA molecules and reveals the salient features of the overall dynamics. This "noise lubricity" may have broader applicability from molecular to macroscopic locomotion.

**TOC GRAPHICS**

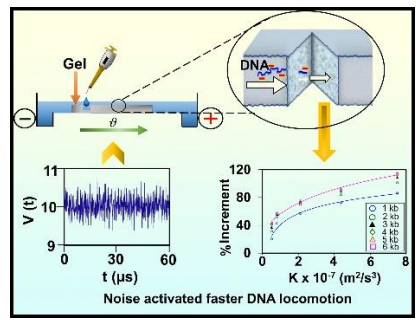

# Introduction

Children learn how to use noise to release a stuck (due to some tiny obstacles) toy car on an inclined plank by tapping it, from their experience. One can also tap on a horizontal platform having a pile of sand to allow free spreading of the sand from the 'jammed' state. The tapping acts as an external perturbation or noise that provides sufficient energy to the system to overcome the surface defects on the plank or the jamming potential barrier in the case of a sand pile. Similar scenarios are encountered almost every day such as the motion of a stuck tiny water droplet either on a windshield of a car on a rainy day or on a shower curtain during a bath. Noise is generated from the random blow of the wind in the former case, whereas, for the later, it is the self-excited oscillations due to the condensation of steam and/or coalescence of multiple droplets.[1] The potential barriers in such cases originate from the solid friction or its surrogate, - contact angle hysteresis, at the solid-solid or solid-liquid interface, respectively. The solid friction restricts the diffusive as well as biased motion of the nano-micro[2] to the large-scale macroscopic objects.[3–5] The friction also affects the dynamics of the charged particles or long-chain molecules like DNA or proteins through a narrow pore.[6,7] Burlatsky and Deutch pointed out that the electrophoretic motion of DNA is hindered by solid friction that originates from the DNA-gel matrix interactions along with the viscous dissipation between DNA and the buffer-solvent that is present inside the pores.[8-10] [also see the responses of Burlatsky and Deutch to the technical comments by Viovy and Duke.[11]] This solid friction is engendered from the adhesive interactions between the pore-wall of the gel matrix and the flexible DNA molecules, rubbing between them, distortion of the gel fibers, and other forms of local kinetic energy dissipation.

To detach from the wall of the gel matrix, an electric field-driven DNA molecule has to overcome these energy dissipative interactions, characterized by critical forces. Constrictions, defined by the dimensions of the pores, govern the solid frictional forces. As the distribution of the pore size spans over a wide range, one can expect a wide distribution of critical forces. These critical forces can be overcome when enough energy is supplied to the DNA molecules. Here we report a novel approach for faster transportation of DNA molecules that ensues from the noise-activated subcritical detachment of DNA from the gel matrix in an electrophoretic setting. The experimental configuration promotes rapid sorting for DNA or protein fingerprinting.

**Materials and methods**

**Materials.** Agarose powder and Ethidium bromide were procured from Loba Chemie. Tris Acetate-EDTA (TAE) buffer was purchased from Himedia, and the DNA 1 kbp to 10 kbp molecular weight (MW) ladder was obtained from Takara Bio Inc. Miniphor UVT System (Serial No. 106888 GB) was used for the horizontal gel electrophoresis and as a DC power supply unit (GeNei Electrophoresis Power Supply, 0-500V, 0-500 mA) was used. For the noise generation and amplification of the signal, an arbitrary waveform generator (SIGLENT SDG- 1062X) and an amplifier (Q44 Keysight 33502A 2-Channel Isolated Power Amplifier) were used.

**Preparation of gel.** 0.8% (w/v) Agarose was prepared by dissolving 0.8 g of agarose powder in 100 ml of 1X Tris Acetate-EDTA (TAE) buffer. A homogenous solution was prepared by first mixing thoroughly and then kept in a microwave oven (Make: Samsung Model No: MC28H5023AKTL) at a power of 900 W for a period of 1 min. The solution temperature was ~ 100 ºC. The solution thus prepared was cool down to ~ 30 °C and subsequently, 3 μL of 0.5 mg/ml aqueous Ethidium bromide (EtBr) was added to the solution. The EtBr facilitates the detection of DNA bands under UV illumination. TAE buffer was used to maintain the pH of the solution at ~ 8. The gel was cast onto a gel electrophoretic platform by pouring the agarose solution into the gel chamber wherein a comb was placed at one end of the gel platform to form the wells into which the DNA would be loading before electrophoresis. The agarose solution was allowed to be set at room temperature for 2 h.

**Estimation of pore size.** The gel was initially frozen at – 20 °C for overnight and then vacuum dried at 20 mTorr and – 103 °C in a Lyophilizer for 24 h before the FESEM analysis. Although sufficient care was taken while measuring the pore size distribution, it is undeniable that the freezing and drying may distort the pores marginally than in the wet state. However, we are interested in the nature of the pore size distribution rather than the absolute pore size. Thus, the freeze-drying technique adopted here is sufficient to provide adequate information about the distribution. From the FESEM image, the pore area was identified using an open source software, ImageJ (see supporting information (SI)). Local contrast-based thresholding was performed to get the area and the distribution of the pore size (area) was obtained from the statistical analysis of the same using Origin Pro 9.0 64 Bit software.

**Instrument.**

Waveform generator (Model: SIGLENT SDG 1062X) was used for generating Gaussian white noise in the form of voltage pulses V(t).

Oscilloscope (Model: RIGOL DS1052E) was used to observe and collect the voltage data (V(t)), generated by the waveform generator for further analysis.

Amplifier (Model: Q44 Keysight 33502A 2-Channel Isolated Amplifier) was used to amplify the Gaussian noise signal generated from the waveform generator.

Electrophoretic chamber (Model: Miniphor UVT System 106888GB) was used for performing the agarose gel electrophoresis for the separation of the DNA bands.

UV illumination/gel documentation chamber ( Model: Chemidoc XRS+ System with Image Lab Software #1708265) was used for visualization of the DNA bands.

Lyophilizer (Model: LABCONCO Freezone 4.5L Benchtop Freeze Dryer, #720401000) was used for freeze drying the agarose gel for FESEM analysis.

FESEM (Model: JSM 7610F, JEOL, JAPAN) was used for imaging the freeze-dried agarose gel section to determine the pore size distribution.

**Experimental procedure.** The DNA 1 kbp to 10 kbp molecular weight ladder was loaded into the wells of the gel. Any bubble formation during the loading of the DNA was avoided carefully. The gel was run at 10 V DC at a constant current of ~ 1 mA across a separation distance of $d = 13$ cm between two Pt electrodes. The duration of the electrophoresis was 4 h. The temperature of the gel and the TAE buffer was maintained at 23±1 °C. Visualization of the DNA fragments was done using a Gel Doc (Chemidoc XRS+) instrument. The system software captures the images at various modes of magnification. Gaussian white noise was generated and amplified using a function waveform generator and a noise amplifier, respectively. The waveform generator and the amplifier were connected in series with a DC voltage source as schematically shown in Fig. 1. To characterize the noise, the output from the waveform generator as voltage signal via an amplifier and an oscilloscope was collected. The stored data was then extracted for further analysis. The minimum interval for the sampling data was 40 ns. The images of the DNA bands after electrophoretic separation were analyzed using the open-source ImageJ software.

## Results and Discussion

**Noise activated DNA translocation.**

The experimental setup was similar to the conventional agarose gel electrophoresis under the influence of a constant bias voltage with a provision for introducing Gaussian noise as time-dependent voltage input. A function generator, along with an amplifier, was attached with a bias DC voltage source in a series connection, and the resultant potential was applied across a gel in an electrophoretic set-up through a pair of Pt electrodes and Tris Acetate-EDTA (TAE) buffer (Figure 1).

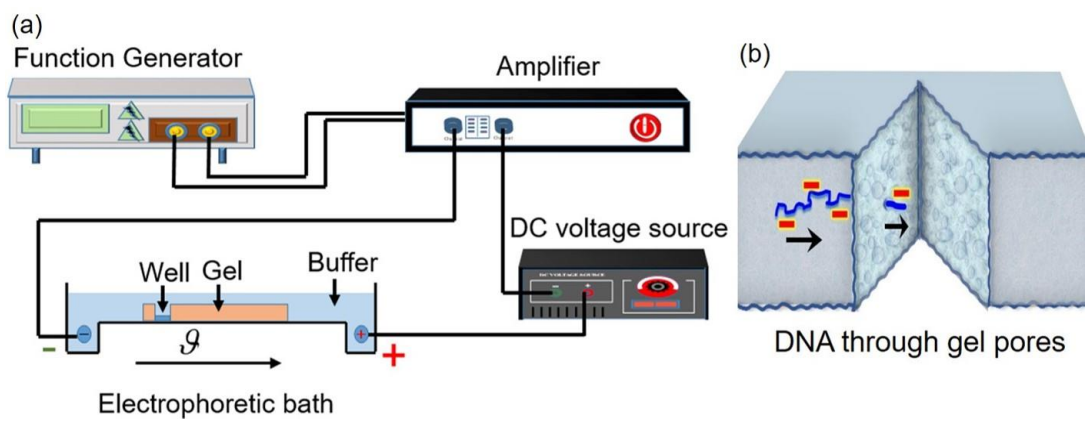

**Figure 1.** Schematic representation of the experimental setup. (a) The function waveform generator is connected with a signal amplifier. The output from the amplifier is connected in series with a constant DC voltage source and the resultant output is applied across the agarose gel in the electrophoretic chamber. The arrow with the symbol denotes the direction of the DNA band motion. (b) The schematic of the enlarged gel cross-section depicts negatively charged DNA experiencing friction while traversing through the pores of the gel matrix.

Under the influence of a bias voltage, the negatively charged DNA fragments slither through the gel having the network of porous tortuous micro/nanotubes, the birth of which itself is engendered by the longitudinal transport of DNA through the gel matrix, similar to the mechanism proposed for entangled polymers.[10,12] Keeping the bias voltage fixed at 10 V, the intensity of the Gaussian noise was controlled by amplifying the voltage amplitude of the noise.

In presence of the noise and with the increase of the noise amplitude, the mobility of the DNA increases and at a given noise, the mobility decreases with the increase of the MW of DNA (Figure 2 a,d).

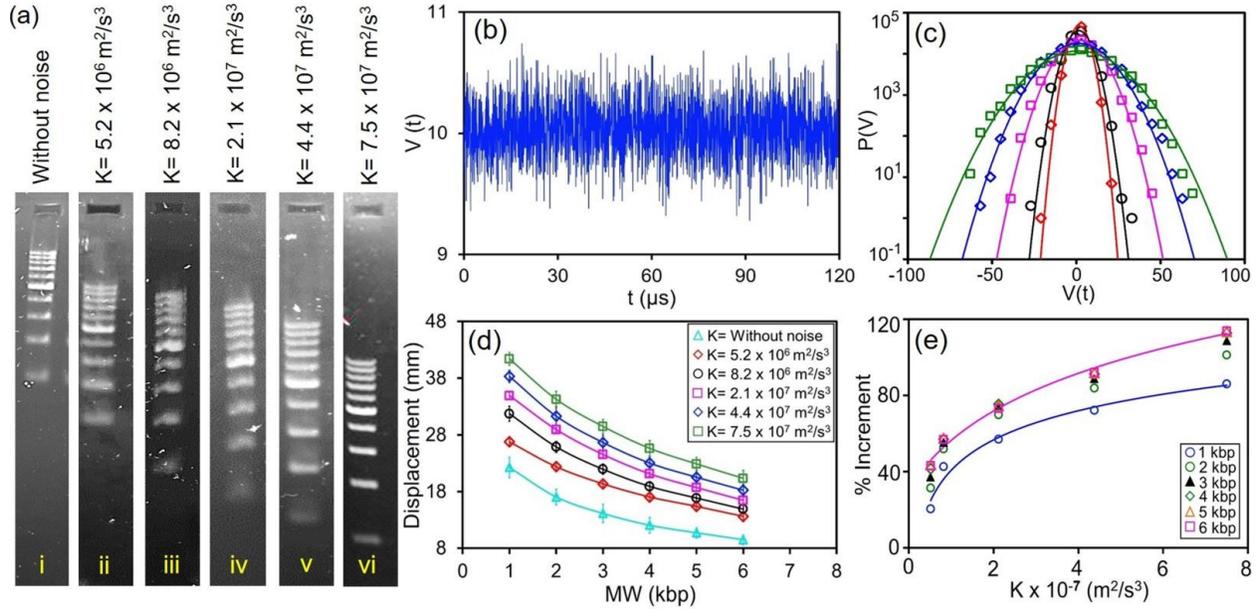

**Figure 2.** Effect of Gaussian noise on electrophoretic separation of DNA. (a) Agarose gel electrophoresis of DNA ladder (1-10 kbp) at 10 V bias for 4 h without any noise (i) and with the noise of different powers (K) depicted on the images (ii-vi). The DNA fragments in the image start from the well (dark rectangle at the top) and then move forward (downward in the image) to the positive electrode during electrophoresis. The band for 10 kbp is the closest to the well and the band for 5 kbp has maximum brightness. (b) A typical example of a time series of noise as voltage input at a low power of the noise. (c) Semi log plot of the input noise depicting Gaussian distribution with mean $\langle V(t) \rangle = 0$. The symbols indicate the noise input corresponding to the power shown in Figure 2d: cyan triangle (△) - without noise, red diamond (◊) – K= 5.2×10$^6$ m$^2$/s$^3$, black circle (○) - K= 8.2×10$^6$ m$^2$/s$^3$, pink square (□) - K= 2.1×10$^7$ m$^2$/s$^3$, blue diamond (◊) - K= 4.4×10$^7$ m$^2$/s$^3$, and green square (□) - K= 7.5×10$^7$ m$^2$/s$^3$, (d) Displacement of the 1 kbp to 6 kbp after 4 h of gel electrophoresis without noise and with noise having different powers shown in the inset. To minimize the error, only estimation of the displacement up to 6 kbp was considered. The error bar represents the standard deviation from the results obtained from 9 sets of experiments. (e) % Increment of the displacement was estimated for 1 – 6 kbp (black circle (○), green circle (○), filled black triangle (▲), green diamond (◊), orange triangle (△), pink square (□) represent 1 to 6 kbp respectively) at different power of the noise input. The percentage increment in displacement was estimated as % Increment= $\frac{(x_n - x_c) \times 100}{x_c}$, here $x_n$ represents displacement for a particular DNA fragment ($n$) at a particular noise, and $x_c$ is the displacement of the corresponding DNA fragment

from the conventional gel electrophoresis without any noise. A typical displacement vs MW of DNA at various power of the noise but at a bias of 50 V is presented in SI (Fig S1).

Under the influence of the noise having power, $K = 5.15 \times 10^6$ m$^2$/s$^3$ (see Supporting Information (SI) for power calculation), almost 20 % enhancement in the mobility of 1kbp DNA over the conventional gel electrophoresis was achieved. The mobility of 1 kbp DNA increased further up to 86 % with the amplified noise having the power of $K = 7.5 \times 10^7$ m$^2$/s$^3$. The increment of mobility was found to be more than 113 % for 6 kbp DNA at the same power of the noise (Figure 2e).

As reported by Burlatsky et al. sorting of the negatively charged DNA fragments based on the molecular weight is only possible in an electrophoretic setting because of the solid friction offered by the gel.[8,13] Otherwise, free electrophoresis only in a buffer, without the gel, shows similar mobility towards a positive electrode for the large DNA molecules (having base pairs larger than ~ 400 bp).[14,15] This suggests that the linear kinematic friction is not sufficient to resolute the large DNA fragments. The resolution effect due to the presence of non-linear solid friction endures even in the presence of noise with increased mobility.

It is the general notion that displacement fluctuation of a free Brownian particle in a thermal bath exhibits Gaussian distribution. This is true when the particle experiences linear kinematic friction where the cause of friction is coupled with the source of the noise, - the heat bath. However, the non-Gaussian asymmetric tail of the displacement distribution, especially at the larger fluctuations, is observed when a particle/object interacts with a surface through Coulombic solid friction [5]. A similar observation is also reported in the case of a colloidal particle diffusing along a lipid bilayer tube or diffusing through the entangled F-actin network.[16] The use of external

Gaussian noise alleviates the effect of the nonlinear solid friction ("Noise-lubricity") and enhances the mobility retaining the resolution characteristics of the gel (Figure 2a). An earlier report suggests a similar subcritical detachment of a soft elastic body from a rigid contactor in presence of mechanical noise that promotes diffusive exploration of different states in an energy landscape and selects the least action pathway.[17]

**Modified Langevin Model.**

The noise used here is the time-dependent accelerations $\gamma(t) = \frac{qV(t)}{md}$, experienced by each base pair unit (see SI). Here $m$ is the average mass of a unit base pair, $q$ is the total charge of a base pair, $d$ is the distance between the two electrodes, and $V(t)$ is the delta correlated time-dependent voltage (Figure 2b). The distribution of the noise pulses is Gaussian (Fig. 2 (c)). This noise-induced drifted motion of the DNA molecules can be approximated by a modified Langevin equation:

$$\frac{d\vartheta}{dt} + \frac{\vartheta(t)}{\tau_L} + \sigma(\vartheta)\Delta(x) = \bar{\gamma} + \gamma(t), \qquad (1)$$

Here, $\bar{\gamma} = \frac{Eq}{m}$ is the bias driving force per unit base pair with $E = \frac{V}{d}$ being the electric field, pertinent to the applied constant bias voltage $V$. $\Delta(x)$ is the space-dependent nonlinear solid friction associated with a signum function, $\sigma(\vartheta) = \frac{\vartheta}{|\vartheta|}$, that defines the direction of the solid friction opposite to that of the instantaneous velocity, $\vartheta$. $\tau_L$ being the Langevin relaxation time, $\frac{\vartheta}{\tau_L}$ accounts for the linear kinematic friction. This Langevin description enables us to capture the salient features of the external noise-activated dynamics of the center of mass of DNA.

Unlike the earlier reports,[18] here we have considered the motion of the center of mass of a DNA molecule. Often, in gels having a high agarose concentration (corresponds to relatively small pore sizes, thus associated with higher average friction), DNA gets trapped inside the constricted pores in loop-like conformations. In this scenario, due to size fluctuations, and the interplay between the relaxation time scale and the time period of the external forcing pulses, DNA exhibits anomalous mobility that results in band inversion.[19-21] The location of the DNA at a given instant of time must be associated with a conformation that corresponds to a particular friction value. This friction value has to be conserved as far as the conformation is maintained and the associated location is occupied by the DNA.

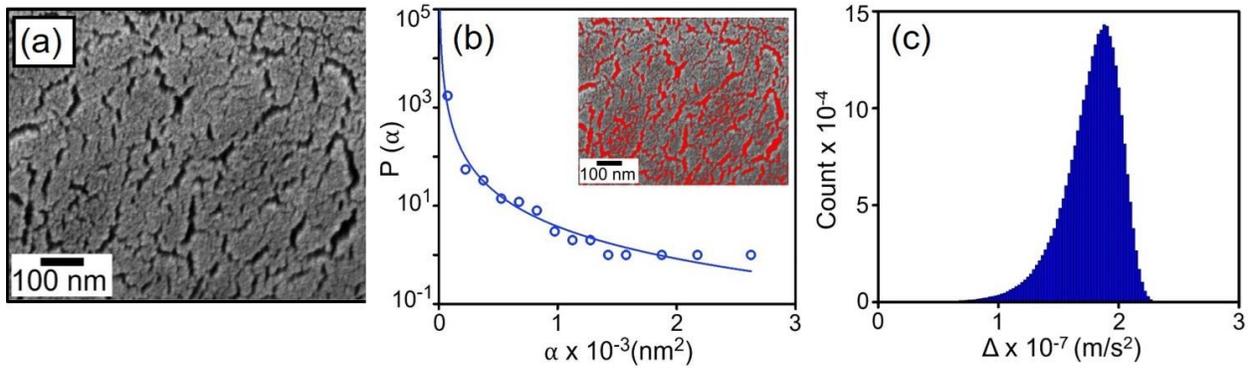

**Figure 3.** Agarose gel pore size and solid friction. (a) SEM image of the 0.8% agarose gel. The gel was initially frozen at – 20 °C for overnight and then vacuum dried at 20 mTorr and – 103 °C in a Lyophilizer for 24 h before the SEM analysis (b) Open-source ImageJ software was used to estimate the pore size (inset image showing the thresholding of the image). The probability distribution function $[P(\alpha)]$ of the pore size ($\alpha$) (open blue circle) is fitted with the allometric regression of the form $[P(\alpha)] \sim \alpha^{-2.2}$. (c) Extreme value distribution of solid friction $\Delta$ used for the numerical simulation of the Langevin Eq. (1).

Apart from this, as evident from the FESEM image of the gel, the pore size ($\alpha$) is distributed over space (Figure 3a). As the solid friction is related to the constriction, defined by the pore walls, the variation of $\Delta$, as a function of the space is thus justified. The allometric decay of the pore size (Figure 3b) readily suggests that the frequency of encounters with the high solid friction sites (small pore) by a DNA molecule is more than the number of encounters with the low solid friction sites (large pore). Thus, one can qualitatively assume an 'Extreme value' distribution of $\Delta$: $P(\Delta) =$

$\frac{1}{s} exp\left(\frac{\Delta-\Delta_m}{s}\right) exp\left[-exp\left(\frac{\Delta-\Delta_m}{s}\right)\right]$, randomly dispersed over the space. Here, $\Delta_m$ is the magnitude of $\Delta$ with the maximum occurrence, and $s$ is the scale parameter of the distribution. A typical $\Delta$ value distribution is depicted in Figure 3c. While the DNA molecules slither through the pores, this random space-dependent $\Delta$, implicitly takes into account the time-dependent molecular conformations. Granick's group reported an interesting observation while imaging a single DNA molecule transporting through an agarose gel.[22] They demonstrated that the trailing end and the leading end of a DNA molecule are stuck at the same position of an agarose matrix for quite some time before leaving the position. This observation straightaway suggests that there is a distribution of constrictions throughout the gel matrix, which encouraged us to consider the random distribution of $\Delta$ as a function of position. Considering this space-dependent $\Delta$, the numerical integration of Langevin Eq. 1, following Gillespie,[23] successfully simulates the trajectories of the DNA and captures the salient features of the dynamics (See Figure S2 in the section S3 of SI). Drift velocity calculated from the simulated trajectories are in reasonable agreement with the experimental results (Figure 4a). For the longer duration of the electrophoresis with a large number of DNA molecules, one can still assume an approximate average $\Delta$ value for a particular gel with a specific agarose concentration. Considering the linear approximation of Eq. (1) and from the Fokker-Planck solution in the velocity space, one can estimate the average drift velocity as[3,24-26]:

$$\vartheta_d = \frac{\bar{\gamma}\tau_L}{1+\frac{\Delta^2 \tau_L}{K}}, \tag{2}$$

This approximate drift velocity ($\vartheta_d$) agrees well with the experimental drift velocity of the DNA molecules with $\Delta \sim n^{0.1}$ and $\tau_L \sim n^{-0.4}$ (Figure S4 in SI).

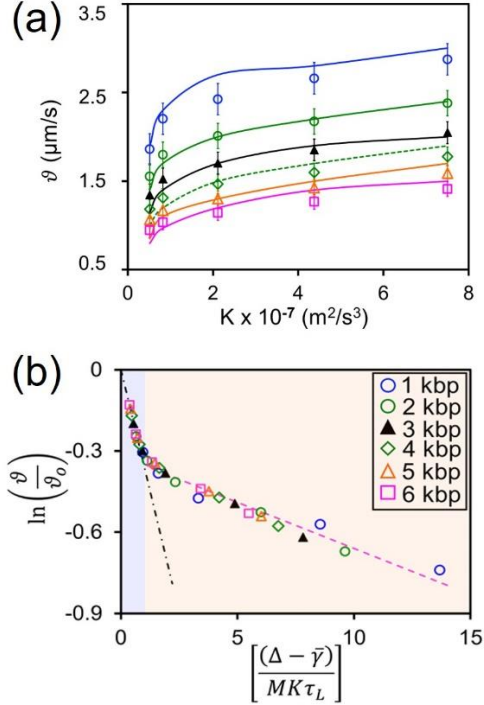

**Figure 4.** (a) Comparison of the drift velocity obtained from the 30 simulations for each case (curves) and the experiments (solid symbols). The error bar for the experimental data depicts the standard deviation of the results obtained from 9 sets of experiments in each case. (b) Arrhenius/Non-Arrhenius-like behavior. The velocity data for all the fragments (1 – 6 kbp) are plotted following Eq. (3) having $\Delta \sim n^{0.1}$ and $\tau_L \sim n^{-0.4}$. The blue shaded (high $K$) region shows noise-activated Arrhenius-like behavior and is represented with a black dash-dot line. Whereas, the yellow shaded (low $K$) region depicts Super-Arrhenius behavior, – the velocity obtained is much higher (shown with a pink dash line) than that expected from Arrhenius prediction (black dash-dot line).

**Escape rate of the DNA molecule.**

As discussed above, the energy barrier, $E_a$, originates from the nonlinear interactions of DNA molecules with the gel matrix. Solid friction, $\Delta$, being the significant contributor to the nonlinear interactions, one can assume the scaling of the energy barrier $E_a \sim \Delta$ as the first approximation. The rate of the detachment of DNA molecules from the gel matrix manifests in the drift velocity of the molecules and can be represented as:

$$\vartheta = \vartheta_o exp\left[-\frac{C(\Delta-\bar{\gamma})}{MK\tau_L}\right], \qquad (3)$$

Here $C$ is a constant and $\vartheta_o$ is the critical velocity while the biased force per unit mass, $\bar{\gamma}$, is sufficient to transcend the energy barrier, i.e. $\bar{\gamma} \approx \Delta$. The energy supplied through the external noise, $MK\tau_L$, represents the mechanical analog to thermal energy $k_B T$, where $M$ is the molecular weight of DNA. While $ln\left(\frac{\vartheta}{\vartheta_o}\right)$ is plotted against $\left[\frac{(\Delta-\bar{\gamma})}{MK\tau_L}\right]$ all the velocity data for different DNA fragments merge into a single master curve with the same average $\Delta$ values used for Eq. (2) (Figure 4b). Although at the high power of the noise ($K$), the velocity follows the Arrhenius-Eyring[24,27] like equation (black dash-dot line in Figure 4b), it exhibits Super-Arrhenius-like behavior (pink dashed line in Figure 4b) at a low power of the noise. Arrhenius expression for a kinetic process implicitly assumes a single and well-defined rate-limiting energy barrier to transcend. However, the free energy landscape of gel electrophoresis is populated with multiple metastable states separated by saddle points. Thus a process in which a particle or molecule maneuvers through these pathways overcoming the multiple saddle points, bypassing the pinnacles of the energy barriers, exhibits Non-Arrhenius kinetics.

The super Arrhenius behavior is often observed in thermally activated viscous slowing down of a weakly bonded glass-forming liquid in a super-cooled regime.[28] At a temperature lower than a characteristic threshold temperature T* for a liquid, its viscosity follows Super-Arrhenius behavior. This is attributed to the collective and cooperative nature of the thermally activated system at low temperature (T<T*) for a weakly bonded system. At the low power of the noise (analogous to low temperature), super Arrhenius behavior of DNA translocation indicates that the dynamics are influenced by the cooperative motion of DNA, affected by prominent non-linear friction. The magnitude of this friction is distributed over a wide range and is the source of space-dependent energy barriers. DNA molecule translocates through the different pathways meandering downhill of the energy landscape towards the global equilibrium. In this quest, DNA may stick to a local energy pit until and unless a high-energy noise pulse rescues it. However, the overall motion of the DNA is collectively emerging as space averaged drift velocity. At a high power of the noise, the effect of the nonlinear friction is alleviated by frequent such rescue events. Thus, free-flowing yet noise-activated Arrhenius characteristics emerge at high athermal energy.

## Conclusions

Our experiments demonstrate that the mobility of DNA molecules in gel electrophoretic settings can be significantly faster (~ 100 % or more) than the conventional gel electrophoresis[29] with the aid of external Gaussian noise. The drift velocity of the DNA induced by activated "noise-lubricity" follows the Arrhenius-Eyring-like escape rate at the high energy of the external noise. Whereas, at low energy, the cooperative dynamics of the DNA molecules impart super Arrhenius-like behavior. A modified Langevin simulation successfully predicts the drift velocity for an applied bias and noise, along with the consideration of space-dependent nonlinear solid friction, originating from the wide distribution of the gel pore size. In contrast to the conventional notion, this study reveals that an enormous amount of solid friction (~ $10^7$ m/s$^2$) is operative at the interface of the DNA-gel matrix. Activated translocation can be observed from the submolecular electronic level (due to thermal noise)[30] to the macroscopic objects (due to mechanical noise)[31] as well. This work suggests strategies for noise-activated faster DNA fingerprinting and sets up the platform for advanced research on resonance-induced super mobility[32] for the isolation of a specific protein fragment from a crowd.

## Associated Content

Supporting Information associated with this article can be found in the online version at: …

The supporting information contains details of materials and methods, calculations for the power of the noise, details of numerical simulation, displacement distribution, drift velocity from the linearization of the Langevin equation, the power spectrum of the experimental input noise, FESEM images of gel before and after the experiments, and steps to estimate the pore size.

## Author Information

### Corresponding Author
**Partho Sarathi Gooh Pattader-** Department of Chemical Engineering, Indian Institute of Technology Guwahati, Assam, 781039, India; Centre for Nanotechnology, Indian Institute of Technology Guwahati, Assam, 781039, India; Jyoti and Bhupat Mehta School of Health Science & Technology, Indian Institute of Technology Guwahati, Assam, 781039, India;Email: psgp@iitg.ac.in

### Authors
**Aniruddha Deb-** Department of Chemical Engineering, Indian Institute of Technology Guwahati, Assam, 781039, India.
**Prerona Gogoi-** Department of Chemical Engineering, Indian Institute of Technology Guwahati, Assam, 781039, India.
**Sunil Kumar Singh-** Department of Chemical Engineering, Indian Institute of Technology Guwahati, Assam, 781039, India.

## Acknowledgment

PSGP thanks DST SERB, Grant no: CRG/2019/000118, ICMR grant no. 5/3/8/20/2019- ITR, and MeitY - grant no. 5(1)/2022-NANO for financial supports. PSGP acknowledges the fruitful discussions with Prof. Manoj K. Chaudhury, Lehigh University, USA. We thank Prof. Siddhartha Sankar Ghosh for letting us use his facility for Gel documentation.

## Data availability
Te datasets generated during the current study are available from the corresponding author on reasonable request.

# Noise Activated Fast Locomotion of DNA through Frictional Landscape of Nanoporous Gel

Aniruddha Deb[1], Prerona Gogoi[1], Sunil K. Singh[1], Partho Sarathi Gooh Pattader[1,2,3]*

[1]Department of Chemical Engineering, Indian Institute of Technology Guwahati; Assam, 781039, India.

[2]Centre for Nanotechnology, Indian Institute of Technology Guwahati; Assam, 781039, India.

[3]School of Health Science & Technology, Indian Institute of Technology Guwahati; Assam, 781039, India.

*Corresponding author. Email: psgp@iitg.ac.in

## S1. Power of the noise

The power of the noise is calculated per base pair of the nucleotide as follows:

Time-dependent random acceleration experienced per base-pair $\gamma(t) = \frac{2CV(t)}{md}$ where

C = Charge of an electron,

d = Distance between two electrodes of the electrophoretic chamber,

m = mass of a base pair,

V(t) = time-dependent voltage generated by function waveform generator,

The mean of the noise input, $\langle \gamma \rangle = 0$,

The power of the noise then calculated as $K = \langle \gamma^2 \rangle dt$,

The $\langle \gamma^2 \rangle$ denotes the mean square acceleration and *dt* is the noise pulse duration (40 ns).

The noise correlation time scale $\tau_c$ can be obtained from the corner frequency of the power spectra of the Gaussian white noise as described later in the section S5. The correlation time obtained by this method is ~ 6 ns. The autocorrelation function of the noise, as shown in the figure

S5 (b), also suggest that the noise is not fully uncorrelated at very small time scale of the order of ns. Thus, the power of the noise is nominally defined as the product of the mean square acceleration and the pulse width of dt ~ 40 ns which is larger than 6 ns.

## S2. Displacement of DNA fragments at different noise strengths with a bias of 50 V

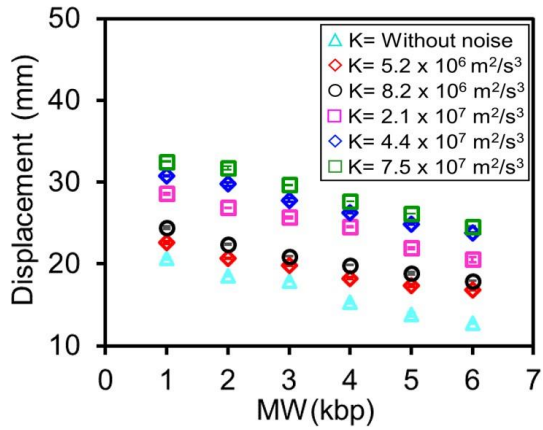

**Figure S1.** Displacement of the 1 kbp to 6 kbp after 90 min of gel electrophoresis without noise and with noise having different powers shown in the inset. Here, the symbols represent the noise powers as follows: cyan triangle (△) - without noise, red diamond (◊) – K= 5.2×10$^6$ m$^2$/s$^3$, black circle (○) - K= 8.2×10$^6$ m$^2$/s$^3$, pink square (□) - K= 2.1×10$^7$ m$^2$/s$^3$, blue diamond (◊) - K= 4.4×10$^7$ m$^2$/s$^3$, and green square (□) - K= 7.5×10$^7$ m$^2$/s$^3$.

## S3. Details of the Numerical Simulation

A numerical simulation of the modified Langevin Equation S1 (Eq. 1 in the main text):

$$\frac{d\vartheta}{dt} + \frac{\vartheta(t)}{\tau_L} + \sigma(\vartheta)\Delta(x) = \bar{\gamma} + \gamma(t) \tag{S1}$$

was carried out following the methodology of Gillespie[1] using MATLAB. The DNA molecule is assumed as a point mass at the center of gravity of the molecule at any point in time. Random noise is generated as a voltage signal using an inbuilt random number generator having Gaussian distribution. The simulations were performed with an integration time step of dt = 0.04 ps so that the ratio of $\frac{\tau_L}{dt} \sim 10$.

**Space dependent Δ**

To introduce the space-dependent Δ, first, a set of random Δ values are generated that follows Extreme value distribution: $P(\Delta) = \frac{1}{s}\exp\left(\frac{\Delta - \Delta_m}{s}\right)\exp\left(-\exp\left(\frac{\Delta - \Delta_m}{s}\right)\right)$. Then each Δ value is assigned to the position, *x*, in such a manner so that over a random length (l) of space, the same Δ value will be experienced. This length (l) also follows a normal distribution.

**Specific conditions for $|\Delta| > |\bar{\gamma} + \gamma(t)|$**

At any instance, if the velocity of the DNA molecule is zero and the $|\Delta| > |\bar{\gamma} + \gamma(t)|$, the velocity will remain as zero unless, at a later time, one pulse of the noise γ(t) will be sufficient enough so that the total acceleration $|\bar{\gamma} + \gamma(t)|$ surpasses the magnitude of the Δ. If at any instance, while $|\Delta| > |\bar{\gamma} + \gamma(t)|$, the velocity is small and less than a critical value $\vartheta_0$, exponential decay of the velocity with the Langevin time scale $\tau_L$ is assumed. Otherwise, the full equation (S1) is integrated numerically.

**Typical Simulated trajectory**

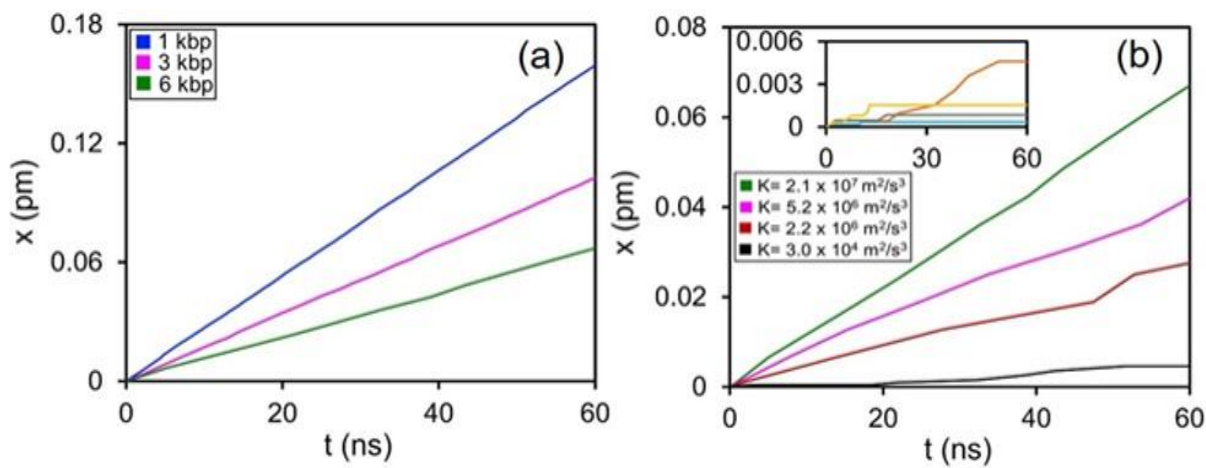

**Figure S2. Simulation results from the Langevin Eq. (S1).** (a) Typical simulated trajectories of the DNA fragments (color code is shown in the inset) in gel electrophoretic setting at a particular noise $K = 2.1 \times 10^7$ m²/s³. (b) Simulated trajectories of 6 kbp DNA fragment at different powers of the noise shown in the figure. Inset shows 4 different trajectories of 6 kbp DNA fragments at low power ($K = 3.0 \times 10^4$ m²/s³) of the noise depicting the randomness and fluctuations during translocation.

Langevin simulation at a very low power of the noise shows considerable 'stick' states in the trajectories of DNA molecules at some high $\Delta$ value (i.e. at a small pore) (Figure S2b inset).

## S4. Displacement distribution

From the trajectory, at different time steps, the displacement jumps were calculated. The probability density function (PDF) of the displacement $\tilde{x}$ are depicted for 1 kbp and 10 kbp in Figure S3A and S3B with a peak shift of $\left(\tilde{x} = \left(x - x_p\right)\right)$. Here $x_p$ denotes the displacement value having maximum count in the distribution of x. Due to the presence of non-linear friction, the distribution is asymmetric and having a non-Gaussian exponential tail. Assuming average solid friction Δ, from the steady-state solution of the Klein-Kramers equation[2,3] the PDF of the velocity distribution can be obtained as:

$$P(\vartheta) = P_o \exp\left(-\frac{\vartheta^2}{K\tau_L} - \frac{2|\vartheta|\Delta}{K} + \frac{2\vartheta Eq}{Km}\right), \quad (S2)$$

The presence of the 2$^{nd}$ term within the argument of the exponential function imparts the asymmetric non-Gaussian tail in the velocity distribution that is evident from the displacement PDF. Simulation with Δ = 0 must give rise to symmetric Gaussian distribution according to equation S1, and thus depicted in Fig S3C for 1 kbp DNA.

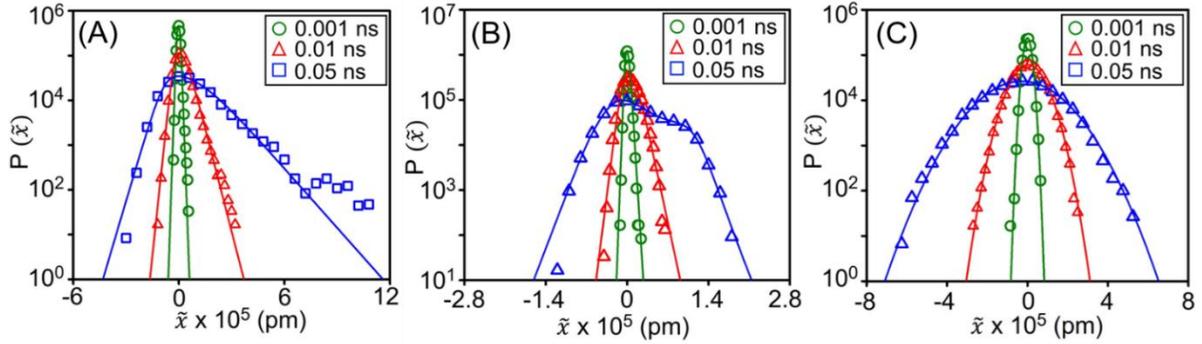

**Figure S3. PDF of displacement fluctuation.** The PDF of the displacement fluctuation in semi-log plot at different time windows (0.001, 0.01, and 0.05 ns) for 1 kbp (A) and 10 kbp (B) with space-dependent Δ. The peak position of the distribution is shifted to zero by $\tilde{x} = (x - x_p)$. The tails of these distributions are asymmetric and exponential. (C) The PDF of the displacement fluctuation for 1 kbp DNA with Δ = 0 shows the Gaussian distribution.

## S5. Drift velocity from the linearization of Langevin equation

From the linearization of the Langevin equation, the approximate drift velocity can be obtained as:

$$\vartheta_d = \frac{\bar{\gamma}\tau_L}{1 + \frac{\Delta^2 \tau_L}{K}}, \quad (S3)$$

Here, $\bar{\gamma} = \frac{Eq}{m}$ is the bias driving force per unit base pair. $\tau_L$ is the Langevin relaxation time and Δ is the average solid friction force per unit base pair. The approximate drift velocity is estimated from the stationary solution of the Fokker Planck equation in velocity space [4,5]. The drift velocity obtained from equation (S3) is shown in Figure S4.

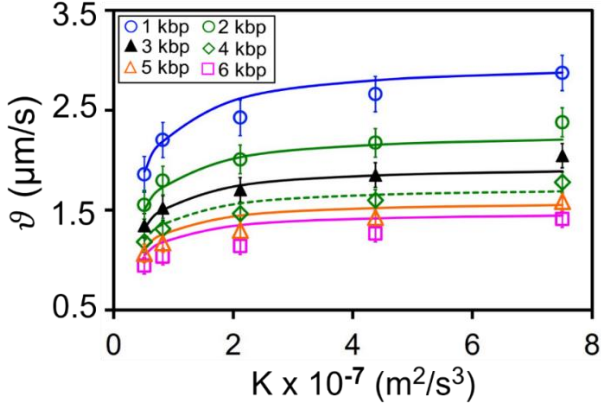

**Figure S4. Drift velocity of the DNA fragments.** Comparison between the experimental drift velocity (symbols) and the velocity predicted by the equation S3 (line) for different DNA fragments.

Although the approximate Eq. (S3) describes the drift velocity reasonably well, its validity is somewhat dubious in the present scenario. From the fitting of the drift velocity using Eq. (S3), the Langevin relaxation time $\tau_L$ is found to be of the order of $\sim 10^{-13}$ s, which is much smaller than the sampling time, dt, of the external noise input $\left(\frac{\tau_L}{dt} \sim 10^{-5}\right)$. Ideally, for Eq. (3) to be applicable, the Langevin relaxation time, $\tau_L$, should be longer than the noise correlation time scale, $\tau_c$. However, the power spectra of the noise reveal that the noise is white till $f_c = 2.5 \times 10^7$ Hz (Figure S5). Considering this $f_c$ as the corner frequency, the approximate correlation time constant, $\tau_c \sim 6$ ns, (from the equation, $\tau_c = \frac{1}{2\pi f_c}$). The estimation of $f_c$ is however limited by the experimental sampling rate of the noise data collection. Thus, one can expect that actual $\tau_c$ may be much smaller than the 6 ns as the $f_c > 2.5 \times 10^7$ Hz. In support of this smaller noise correlation, numerical simulation of the Langevin Eq. (S1), satisfactorily agrees with the experimental drift velocity with $\frac{\tau_L}{dt} \sim 10$ (Figure 4a).

## S6. The power spectrum of the experimental input noise

The power spectrum of the input noise is shown in Fig. S5. The power spectrum of the noise, having total bandwidth (-$f_{max}$ to +$f_{max}$, $f_{max}$ being the maximum frequency) of 50 MHz, is reasonably flat indicating the noise can be considered as white noise up to the maximum frequency $f_{max}$. However, this is limited by the data sampling frequency with dt = 40 ns. The actual bandwidth of the input noise might be larger than this 50 MHz.

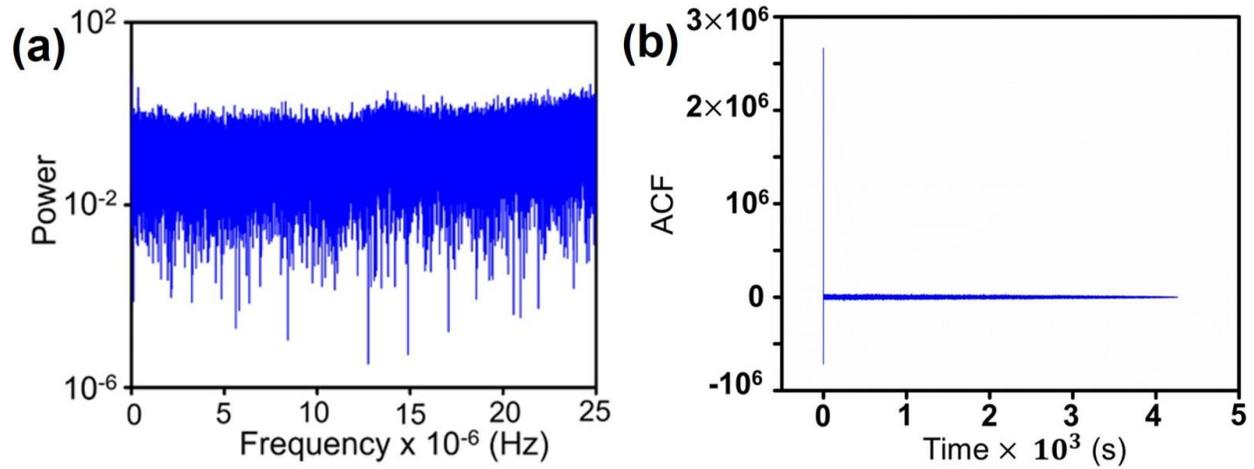

**Figure S5. Power Spectra and autocorrelation of the input noise.** (a) A typical power spectrum of the time-dependent voltage V(t) as the noise input. The spectrum shows that the spectrum is reasonably flat up to the frequency $f_c = 2.5 \times 10^7 \, Hz$. As this estimation is limited by the sampling time dt = 40 ns one can expect the noise is white over a larger frequency domain than $2.5 \times 10^7 \, Hz$. (b) The autocorrelation function (ACF) of the input noise suggests the noise is uncorrelated except at a very small time scale of the order of ns.

## S7. FESEM image of gel before and after electrophoresis experiment with the noise of power $K=7.5 \times 10^7$ m$^2$/s$^3$

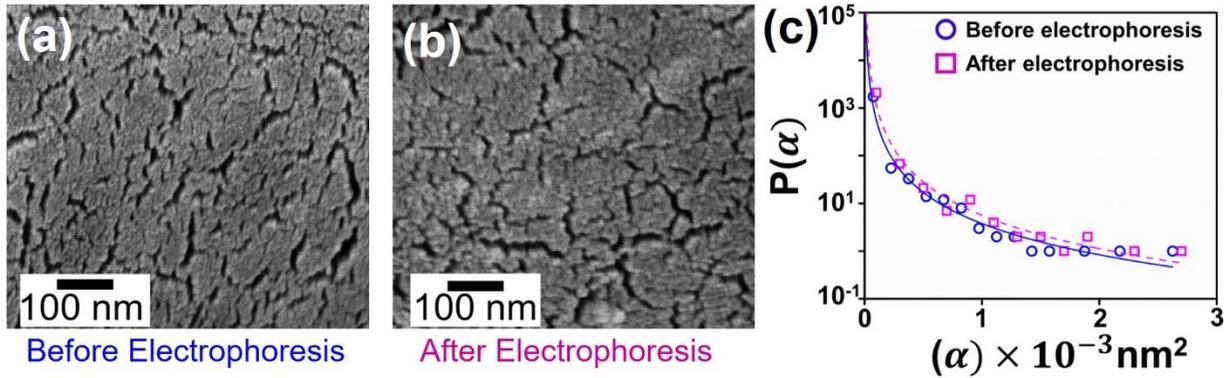

**Figure S6. FESEM image of gel**. FESEM images of agarose gel before (a) and after (b) electrophoresis experiment with noise having power $K=7.5 \times 10^7$ m$^2$/s$^3$. The identical procedure (as mentioned in the main text) of the freeze-drying technique was used before performing the FESEM. The pore size distributions show a negligible difference before and after electrophoresis (c).

## S8. Pore size estimation using ImageJ software

The following steps were followed to estimate the pore size:

1. Initially, the FESEM image was loaded in ImageJ software (Version 1.52s 64-bit).
2. Under the "Analyze" and under the drop down menu, "Set Scale" was selected and from the known distance, the global scale was assigned.
3. From the "Image" option, under the drop down menu the "Adjust" option was selected and thresholding was applied to the pores selecting "red color".
4. Next, from the "Analyze" tab, "Analyze particles" was selected from the drop down menu to get the pore areas.
5. The pore area values were exported in an excel sheet and then transferred to Origin Software (Version 9.0 64 Bit) to get the pore size distribution with proper fitting.